\documentclass[aps,prl,twocolumn,superscriptaddress,groupedaddress]{revtex4}
\usepackage{graphicx}  
\usepackage{dcolumn}   
\usepackage{bm}        
\usepackage{amssymb}   
\usepackage[resetlabels,labeled]{multibib}
\newcites{S}{Supplemental References}
\newcites{t}{test}

\makeatletter

\newcommand{\Rmnum}[1]{\expandafter\@slowromancap\romannumeral #1@}
\makeatother

\makeatletter
\renewcommand{\section}{\@startsection
  {section}%
  {1}%
  {0mm}%
  {-\baselineskip}%
  {0.5\baselineskip}%
  {\normalfont\bfseries}} 
\makeatother

\begin{document}



\title{Coulomb drag in graphene/hBN/graphene moir\'{e} heterostructures}

\author{Yueyang Wang}
\author{Hongxia Xue}
\affiliation{Department of Physics and HK Institute of Quantum Science $\&$ Technology, The University of Hong Kong, Pokfulam Road, Hong Kong, China}

\author{Xiong Wang}
\affiliation{Department of Physics, The University of Hong Kong, Pokfulam Road, Hong Kong, China}

\author{Kenji Watanabe}
\affiliation{Research Center for Electronic and Optical Materials, National Institute for Materials Science, 1-1 Namiki, Tsukuba 305-0044, Japan}

\author{Takashi Taniguchi}
\affiliation{Research Center for Materials Nanoarchitectonics, National Institute for Materials Science,  1-1 Namiki, Tsukuba 305-0044, Japan}

\author{Dong-Keun Ki}
\email{dkki@hku.hk}
\affiliation{Department of Physics and HK Institute of Quantum Science $\&$ Technology, The University of Hong Kong, Pokfulam Road, Hong Kong, China}

\date{\today}

\begin{abstract}
We report on the observation of Coulomb drag between graphene-hexagonal boron nitride (hBN) moir\'{e} heterostructure with a moir\'{e} wavelength of $\sim$14 nm and an intrinsic graphene with a lattice constant of $\sim$0.25 nm. By tuning carrier densities of each graphene layer independently, we find that the charge carriers in moir\'{e} mini-bands, i.e., near the satellite Dirac point (sDP), can be coupled with the massless Fermions near the original Dirac point (oDP), strongly enough to generate a finite drag resistivity. At high temperature ($T$) and large density ($n$), the drag resistivities near both oDP and sDP follow a typical $n^{-\alpha}$ ($\alpha=1.3\sim1.7$) and $T^2$ power law dependence as expected for the momentum transfer process and it also satisfies the layer reciprocity. In contrast, at low $T$, the layer reciprocity is broken in both oDP-oDP and sDP-oDP coupled regions that suggest dominant energy drag. Furthermore, quantitatively, the drag resistivities near sDPs are smaller than those near oDP and they deviate from $T^2$ dependence below $\sim$100 K. These results suggest that the coupling between the carriers in moir\'{e} mini-bands and those in original Dirac bands may not be of a simple Fermi liquid nature.
\end{abstract}

\maketitle

Lattice mismatch and rotational alignment between graphene and hexagonal boron-nitride (hBN) can generate highly tunable moir\'{e} patterns with a wavelength reaching up to $L_M\approx14$ nm which is much larger than the lattice constant of graphene, $a_G\approx0.25$ nm~\cite{yankowitz2012emergence,ponomarenko2013cloning,dean2013hofstadter,hunt2013massive}. Due to such a large periodicity, the moir\'{e} patterns induce superlattice potentials in graphene and create moir\'{e} mini-bands that exhibit distinct electronic properties different from the pristine graphene. Examples include, but not limited to, the emergence of satellite Dirac point (sDP)~\cite{yankowitz2012emergence,ponomarenko2013cloning,dean2013hofstadter,hunt2013massive}, Hofstadter butterfly effect~\cite{ponomarenko2013cloning,dean2013hofstadter,hunt2013massive}, topological valley current~\cite{endo2019topological}, resonant tunnelling~\cite{mishchenko2014twist}, quantum anomalous Hall effect~\cite{serlin2020intrinsic} and ferromagnetism~\cite{chen2020tunable}. These studies demonstrate that the graphene-hBN moir\'{e} heterostructures provide an excellent platform to study the effect of the moir\'{e} mini-bands and their interactions in low-dimensional systems.

\begin{figure*}[t]
  \centering
  \includegraphics[width=0.95\textwidth]{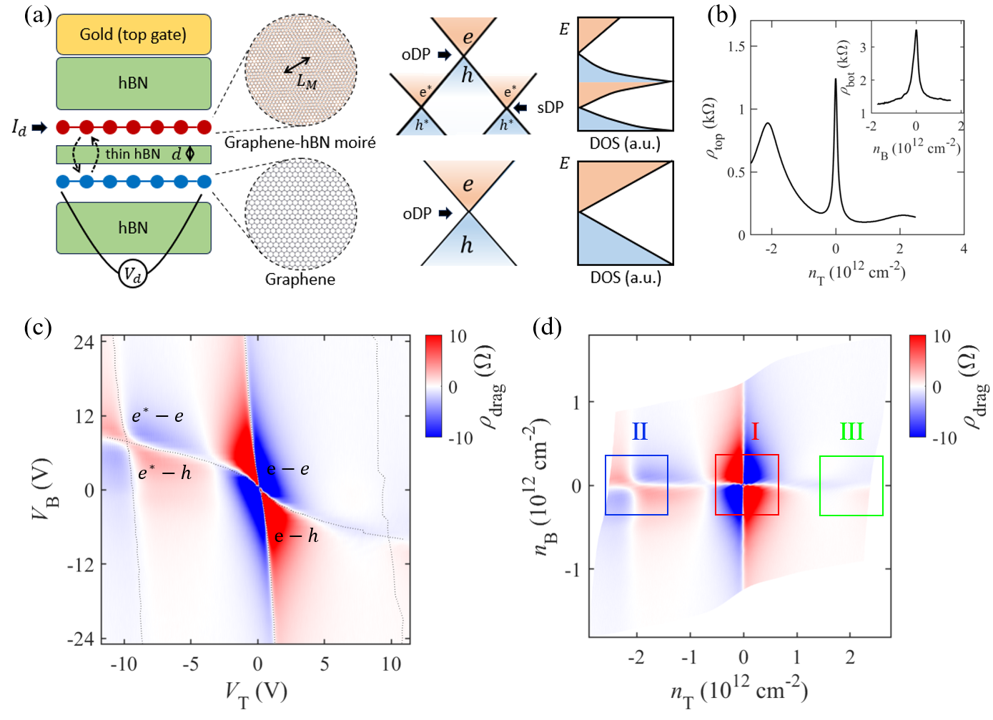}
  \caption{(a) Left: schematics of a typical device structure with top graphene aligned with the hBN to form moir\'{e} patterns and of a measurement configuration for a top drive. Right: schematic illustration of the band structure and corresponding density of states (DOS) in aligned graphene with hBN (top) and in pristine graphene (bottom). The oDP and sDP refer to the original and satellite Dirac points, respectively. Additional hole-like (electron-like) carriers near the sDP are denoted by $h^{\ast}$ ($e^{\ast}$). (b) The longitudinal resistivity of the top graphene layer, $\rho_{top}=R_{top}\times W/L$ ($W$: device width, $L$: length), as a function of $n_{_T}$ exhibiting resistivity peaks at finite densities originating from the moir\'{e} mini-bands with $L_M\approx14$ nm at a twist angle of $\sim$0$^{\circ}$. The inset shows a typical resistivity peak of the pristine graphene for the bottom layer, $\rho_{bot}(n_B)=R_{bot}\times W/L$. (c) A color-scale plot of $\rho_{drag}$ in top and back gate voltages ($V_T$ and $V_B$) measured in a top drive configuration. The dotted lines indicate the positions of the oDPs and sDPs obtained from the local resistivity measurements of the individual layers. (d) $\rho_{drag}$ as a function of $n_{_T}$ and $n_{_B}$ replotted from (c). Regions~\Rmnum{1}, \Rmnum{2}, and \Rmnum{3} are marked by red, blue and green boxes respectively. All data were measured at 210 K.}
  \label{fig:figure1}
\end{figure*}

Coulomb drag experiment, on the other hand, is a direct measure of electron-electron interactions between two different systems that otherwise cannot be coupled with~\cite{solomon1989new,PhysRevLett.66.1216,RevModPhys.88.025003}. During the experiment, a drive current ($I_{d}$) is applied to one of the layers (an active layer) while a voltage ($V_{d}$) is measured in another (a drag or passive layer) that is spatially separated by a thin insulating spacer (see, e.g., Fig.~\ref{fig:figure1}(a)). Since the two layers are electrically isolated, one can observe a finite drag resistivity $\rho_{drag}=rV_{d}/I_{d}$ ($r$: a device width and length ratio), only when the charge carriers in the two layers are strongly coupled by Coulomb interactions~\cite{PhysRevLett.66.1216,tse2007theory,RevModPhys.88.025003}. The relevant studies have indeed shown that Coulomb drag measurements can reveal the nature of interactions between different systems, such as graphene and GaAs 2D electron gas (2DEG)~\cite{g-2dgamucci2014anomalous}, graphene and 2D superconductors~\cite{g-supconductorguo2022longitudinal,g-supconductortao2023josephson}, 1D and 2D systems~\cite{g-NWmitra2020anomalous,g-CNTanderson2021coulomb}, and more.

In this Letter, we have measured Coulomb drag in graphene-hBN-graphene heterostructures where the top (or bottom) graphene is aligned with a thin hBN spacer ($5\sim12$ nm) to form moir\'{e} patterns (see Figs.~\ref{fig:figure1}(a,b)). By tuning charge densities of the top and bottom layers, $n_T$ and $n_B$, independently, we explored the drag effect in two different regions, the region~\Rmnum{1} where both layers are near the original Dirac point (oDP) and the regions~\Rmnum{2} and~\Rmnum{3} where the layer aligned with hBN was doped to reach the sDP in the hole and electron side respectively (see Figs.~\ref{fig:figure1}(c,d)). Interestingly, we find that $\rho_{drag}$ is finite in all regions even when $L_M\approx14$ nm is an order of magnitude larger than the $a_G$. This indicates that the carriers in the moir\'{e} mini-bands filling a large moir\'{e} lattice are coupled with massless Fermions in original graphene bands well enough to induce the drag effect. Moreover, we have found that at high temperature ($T$) and large density, $\rho_{drag}$ follows a typical frictional drag behavior, $\rho_{drag}\propto n_B^{-\alpha}$ ($\alpha=1.3\sim1.7$) and $\propto T^2$, and satisfies the layer reciprocity relation, while below $\sim$100 K, the layer reciprocity is broken and $\rho_{drag}$ in region~\Rmnum{2} no longer follows the $T^2$ dependence. This suggests that at low $T$, the coupling between the carriers in moir\'{e} mini-bands and the original massless Fermions may not be fully described by a simple Fermi liquid momentum transfer mechanism~\cite{PhysRevLett.66.1216,momentsivan1992coupled,momen-gsong2013coulomb}. 

The graphene-hBN-graphene moir\'{e} heterostructures used in this study were stacked by picking up multiple graphene and hBN layers alternatively~\cite{stackwang2013one} during which we aligned at least one of the graphene layers with hBN to form moir\'{e} patterns. The left panel of Fig.~\ref{fig:figure1}(a) shows a typical device structure with the top layer aligned with hBN, consisting of a top gold electrode as a top gate ($V_T$) to tune the $n_T$ and a highly-doped silicon substrate (not shown) as a back gate ($V_B$) to vary the $n_B$ independently (a classical Hall effect was used to estimate $n_{T,B}$ from $V_{T,B}$). Standard e-beam lithography, lift-off, and etching techniques were used to make a Hall bar geometry whose details can be found elsewhere (Fig. S1(a); see Supplemental Materials)~\cite{rao2023ballistic}. Fabricated devices were measured in cryogenic systems equipped with a superconducting magnet using a low-frequency lock-in technique in a current bias (1 $\mu$A, 17 Hz). To measure the $\rho_{drag}$, we drive $I_d$ through the top (bottom) graphene layer and measure a drag voltage $V_d$ at the bottom (top) layer as depicted in Fig.~\ref{fig:figure1}(a) for a top drive configuration. While measuring $V_d$, we have also simultaneously measured the local voltage drop across the drive layer parallel to $I_d$, $V_{top}=R_{top}\times I_d$ (or $V_{bot}=R_{bot}\times I_d$), to identify the position(s) of the DP(s) by synchronizing two lock-in amplifiers (see Fig. S1(b)). As shown in Fig.~\ref{fig:figure1}(b), the local resistivity measurement confirms the presence of the moir\'{e} potential in the top graphene layer with $L_M\approx14$ nm at a nearly zero twist angle and its absence in the bottom layer as intended. 

\begin{figure*}[t]
  \centering
  \includegraphics[width=0.95\textwidth]{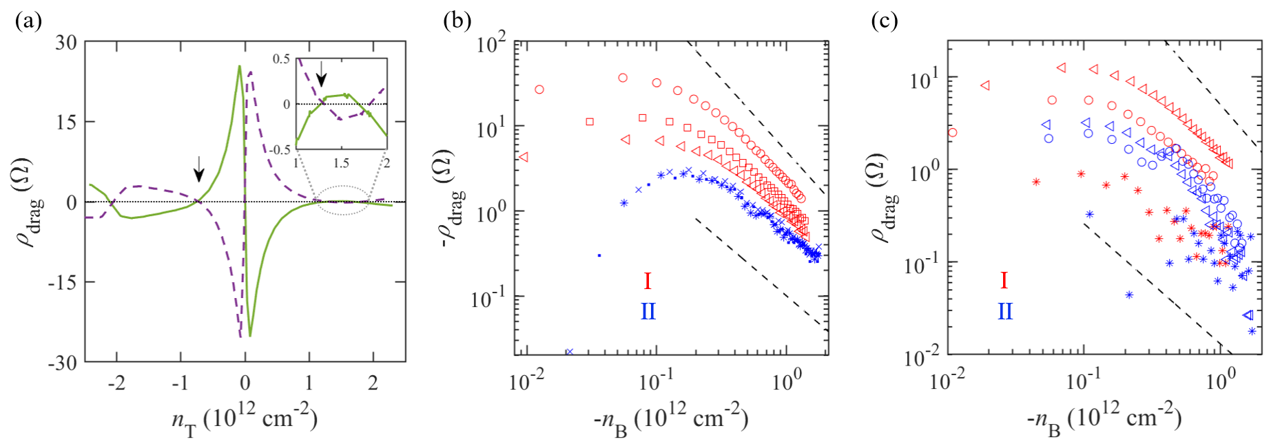}
  \caption{(a) Horizontal line cuts, $\rho_{drag}(n_T)$, from Fig.~\ref{fig:figure1}(d) at $n_B=\pm2.2\times10^{11}$ cm$^{-2}$ (solid green and broken purple lines respectively). The inset magnifies the circled area. The sign changes across DPs and van Hove singularity points (marked by the arrows) are clearly visible. (b) A double-log plot of $-\rho_{drag}(-n_B)$ taken at finite $n_T=n_T^{\ast}= -1.2\times10^{11}$ cm$^{-2}$ (circle and asterisk), $-3.0\times10^{11}$ cm$^{-2}$ (square and cross), and $-4.0\times10^{11}$ cm$^{-2}$ (triangle and dot). (c) A double-log plot of $\rho_{drag}(-n_B)$ at $n_T=n_T^{\ast}=3.0\times10^{11}$ cm$^{-2}$ measured at 210 K (triangle), 140 K (circle), and 50 K (asterisk). In (b) and (c), red and blue colors denote the data from the region~\Rmnum{1} and~\Rmnum{2}, respectively, and the upper and lower dashed lines scale $n_B^{-1.7}$ and $n_B^{-1.3}$.}
  \label{fig:figure2}
\end{figure*}

The drag resistivity measured at 210 K, Figs.~\ref{fig:figure1}(c,d), reveals clear signatures of a finite Coulomb drag in all regions not only near the oDP (region~\Rmnum{1}) but also near the sDPs (regions~\Rmnum{2} and \Rmnum{3}). As shown in Fig.~\ref{fig:figure1}(c), in all regions, $\rho_{drag}$ stays positive when the two layers have opposite carrier types and becomes negative when their carrier types are the same (see Fig.~\ref{fig:figure2}(a) for a clearer sign change in the region~\Rmnum{3}). Interestingly, in addition to the DPs, $\rho_{drag}$ changes its sign at finite densities between the region~\Rmnum{1} and \Rmnum{2} (\Rmnum{3}). This corresponds to the van Hove singularity points where the density of states (DOS) becomes maximum and the carrier type is switched due to the existence of the moir\'{e} mini-bands~\cite{wu2016multiple}. These additional sign changes are more clearly visible in Fig.~\ref{fig:figure2}(a) where the line cuts of Fig.~\ref{fig:figure1}(d) at $n_B=\pm2.2\times10^{11}$ cm$^{-2}$, $\rho_{drag}(n_T)$, are plotted. These sign changes are in good agreement with the frictional momentum drag~\cite{gorbachev2012strong, PhysRevB.83.161401}, confirming that the observed signals are from the Coulomb drag not from the direct tunneling~\cite{tunnelspielman2000resonantly}.

To compare the drag effect in the regions~\Rmnum{1} and \Rmnum{2} more directly, we define $n_T^{\ast}$ as the carrier density of the top layer relative to the sDP and plot $\rho_{drag}$ as a function of $n_B$ at the same values of $n_T$ (the region~\Rmnum{1}) and $n_T^{\ast}$ (the region~\Rmnum{2}). The results for the hole sides in both layers are shown in Figs.~\ref{fig:figure2}(b,c) that exhibit a similar power law dependence of $\rho_{drag}$ in density $\rho_{drag}\propto n_B^{-1.3\sim-1.7}$ (broken lines) at large $n_B$ for both regions as expected for the frictional momentum drag~\cite{narozhny2012coulomb}. Furthermore, as $n_B$ decreases, the $\rho_{drag}$ saturates and even decreases as found in the previous studies~\cite{gorbachev2012strong}. This can be attributed to small Fermi energy at low density compared with the measurement temperature (210 K). Similar behavior is also found in other electron-electron and electron-hole sides (Fig. S2).

Although $\rho_{drag}$ behaves similarly in $n_B$ in both regions~\Rmnum{1} and~\Rmnum{2}, we find that its absolute values and their dependences in $n_T^{\ast}$ are different. First, as shown in Figs.~\ref{fig:figure1}(c,d) and~\ref{fig:figure2}(a,b), the $\rho_{drag}$ is smaller in the region~\Rmnum{2} than in the region~\Rmnum{1}. Secondly, Fig.~\ref{fig:figure2}(b) shows that the $\rho_{drag}$ depends very weakly on $n_T^{\ast}$ in the region~\Rmnum{2} compared with its dependence on $n_T$ in the region~\Rmnum{1}. This small $\rho_{drag}$ and its weak dependence on $n_T^{\ast}$ in the region~\Rmnum{2}, as well as the very weak drag signal in the region~\Rmnum{3} (see, e.g., Fig.~\ref{fig:figure2}(a)), can be qualitatively understood by considering the two intrinsic properties of the graphene-hBN moir\'{e} potential: 1) the large difference in the reciprocal lattices of the moir\'{e} mini-bands and of the original Dirac bands due to an order of magnitude difference in the corresponding lattice constants ($L_M\approx14$ nm and $a_G\approx0.25$ nm) and 2) an electron-hole asymmetry originating from the asymmetric on-site potential of hBN layer and the next-nearest neighbour hoping~\cite{yankowitz2012emergence,wang2015evidence,g-hbnli2018charge}. First, due to the large reciprocal lattice mismatch between the drive and drag layers in the regions~\Rmnum{2} and \Rmnum{3} compared to the region~\Rmnum{1}, a larger momentum transfer $q$ is needed to couple the carriers in these regions than those in the region~\Rmnum{1}, leading to the weaker Coulomb interaction (i.e., smaller $\rho_{drag}$) as the interlayer Coulomb potential exponentially decays in $q$, $V=(2\pi e^2/q)e^{-qd}$ ($d$: an interlayer spacing)~\cite{narozhny2012coulomb,spacingdcarrega2012theory}. Secondly, the electron-hole asymmetry in moir\'{e} potential gives rise to a larger DOS at the sDP in the hole side (the region~\Rmnum{3}) than that in the electron side (the region~\Rmnum{2}), thereby suppressing the drag effect further. These findings illustrate that the Coulomb drag experiment can be used to explore the nature of the interlayer Coulomb interactions in moir\'{e} systems.

\begin{figure}[t]
  \centering
  \includegraphics[width=0.49\textwidth]{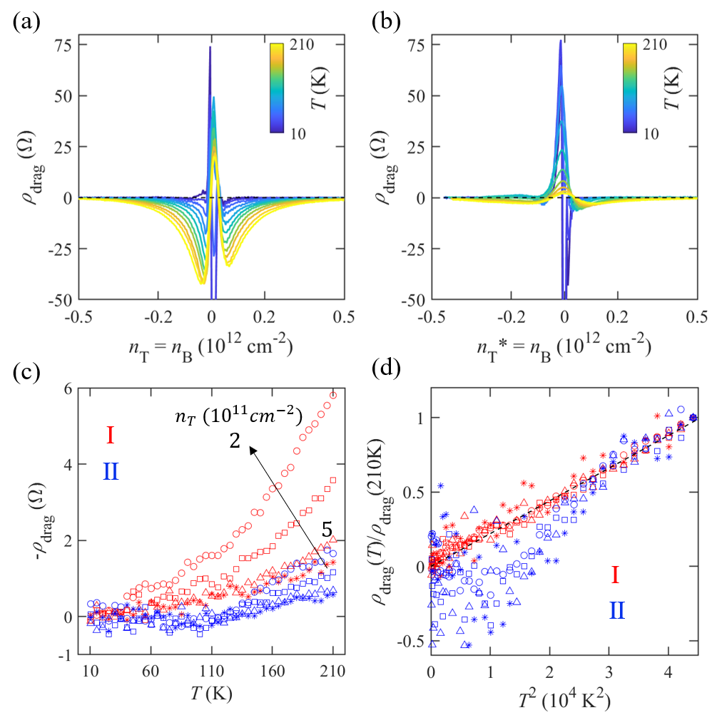}
  \caption{(a,b) $\rho_{drag}$ as a function of $n_T=n_B$ (a), i.e., in the region~\Rmnum{1}, and $n_T^{\ast}=n_B$ (b), i.e., in the region~\Rmnum{2}, at different temperatures from 210 K to 10 K (yellow to blue). (c) $-\rho_{drag}(T)$ taken from (a,b) at finite densities, $2.0\times10^{11}$ cm$^{-2}$ (circle), $3.0\times10^{11}$ cm$^{-2}$ (square), $4.0\times10^{11}$ cm$^{-2}$ (triangle), and $5.0\times10^{11}$ cm$^{-2}$ (asterisk). (d) Normalised $\rho_{drag}(T)/\rho_{drag}$(210 K) as a function of $T^2$. The dashed line indicates the linear fitting. In (c) and (d), red and blue colors denote the regions~\Rmnum{1} and \Rmnum{2}, respectively.}
  \label{fig:figure3}
\end{figure}

Temperature dependence measurements further distinguish the drag behavior in the regions~\Rmnum{1} and~\Rmnum{2}. As shown in Fig.~\ref{fig:figure2}(c), $\rho_{drag}(n_B)$ exhibits stronger temperature dependence in the region~\Rmnum{1} than in the region~\Rmnum{2}. This is more clearly visible in Figs.~\ref{fig:figure3}(a,b) which compare $\rho_{drag}(n_T=n_B)$ from the region~\Rmnum{1} and $\rho_{drag}(n_T^{\ast}=n_B)$ from the region~\Rmnum{2} measured at different temperatures from 210 K to 10 K. The figures show that away from oDP and sDP, $\rho_{drag}$ varies with temperature more in the region~\Rmnum{1} than in the region~\Rmnum{2}. To compare them in detail, we also plot $-\rho_{drag}(T)$ and normalized $\rho_{drag}(T)/\rho_{drag}$(210 K) as a function of $T^2$ (expected for the Fermi liquid~\cite{tse2007theory,narozhny2012coulomb,spacingdcarrega2012theory,hwang2011coulomb,schutt2013coulomb}) at few different values of $n_T$ and $n_T^{\ast}$ in Figs.~\ref{fig:figure3}(c) and (d), respectively. The figures clearly show much weaker temperature dependence of the $\rho_{drag}$ in the region~\Rmnum{2} than in the region~\Rmnum{1} (Fig.~\ref{fig:figure3}(c)) and clear $T^2$ dependence in the region~\Rmnum{1} for all temperature range investigated that contrasts with the non-$T^2$ dependence of the $\rho_{drag}$ in the region~\Rmnum{2} below $\sim$100 K (Fig.~\ref{fig:figure3}(d)). Such a deviation from the $T^2$ dependence at low temperatures suggests that the coupling between sDP and oDP may not be of a simple Fermi liquid nature at low temperatures.

\begin{figure}[t]
  \centering
  \includegraphics[width=0.49\textwidth]{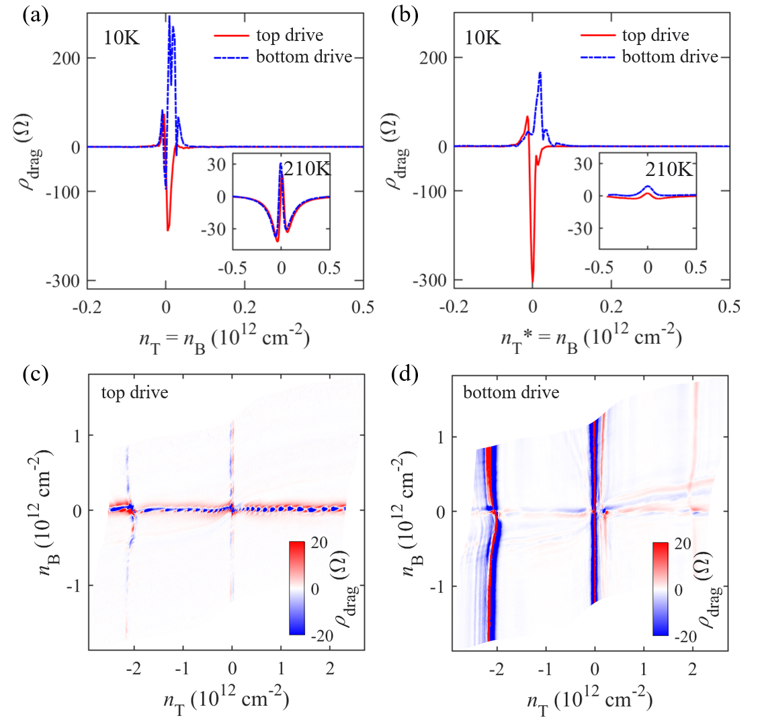}
  \caption{(a,b) $\rho_{drag}$ as a function of $n_T=n_B$ (a) and $n_T^{\ast}=n_B$ (b) measured at 10K. Insets show the data obtained at 210 K. Red solid and blue broken lines indicate the data taken in a top and bottom drive configuration, respectively. (c,d) A color-scale plot of $\rho_{drag}$ in $n_T$ and $n_B$ measured at 5K in top (c) and bottom (d) drive configuration.}
  \label{fig:figure4}
\end{figure}

At low temperatures and near zero density below $1.0\sim4.0\times10^{10}$ cm$^{-2}$, Coulomb drag in both regions~\Rmnum{1} and~\Rmnum{2} exhibit unconventional behavior. As shown in Figs.~\ref{fig:figure3}(a,b), the $\rho_{drag}$ near zero density starts exhibiting a rapid sign change in a small density range as temperature goes below $\sim60$ K. This is more clearly visible in Figs.~\ref{fig:figure4}(a,b) where $\rho_{drag}(n_T=n_B)$ and $\rho_{drag}(n_T^{\ast}=n_B)$ measured at 10 K are plotted in a full scale. Such a sign change is absent at higher temperatures as shown in the insets of Figs.~\ref{fig:figure4}(a,b) and it strongly depends on the drive layer. This indicates that the sharp change of $\rho_{drag}$ near zero density is related to the breakdown of the layer reciprocity at low temperatures. To test this, we measured $\rho_{drag}$ in $n_T$ and $n_B$ at 5 K (Figs.~\ref{fig:figure4}(c,d)) and found that the drag effect nearly only depends on the drag layer. In a top drive configuration, for instance, the $\rho_{drag}$ exhibits strong change in $n_B$ not in $n_T$ (Fig.~\ref{fig:figure4}(c)) while when the bottom layer is driven, the behavior is opposite (Fig.~\ref{fig:figure4}(d)). As found in the previous study~\cite{lee2016giant}, this can be from the energy drag that becomes stronger at lower temperatures. Moreover, near DPs, there are electron-hole puddles in both layers that can be correlated with each other by Coulomb interactions which can give rise to a rapid sign change in a small density range~\cite{gorbachev2012strong,ho2018theory} as found in our experiment (Fig.~\ref{fig:figure4}). From the local resistivity measurements, we estimated the density inhomogeneities of both layers to be in the order of 10$^{10}$ cm$^2$ that roughly matches with the density range where $\rho_{drag}$ exhibits sign changes. This can also explain why there are nearly no signatures of the sign change at the van Hove singularity points in Figs.~\ref{fig:figure4}(c,d) as near the van Hove singularity, the effects of such electron-hole puddles can be suppressed by the large DOS.

In summary, by building graphene-thin hBN-graphene moir\'{e} heterostructures, we have successfully demonstrated that the carriers in the moir\'{e} mini-bands can be Coulomb-coupled with those in an original Dirac band strongly enough to show the finite drag resistivity, despite a fact that their reciprocal lattices are largely different. At high temperatures and large density, we find that the $\rho_{drag}$ in all regions follows a typical frictional momentum drag behavior: a sign change with respect to the carrier types, a power law dependence in density and temperature, and the layer reciprocity. In contrast, at low temperatures, the $\rho_{drag}$ near zero density exhibits a strong dependence in density and depends nearly only on the change in the drag layer. It indicates the strong effect of electron-hole puddles near zero density and the crossover from the momentum to energy drag as lowering temperature that calls for further studies. Nonetheless, our study proves that Coulomb drag experiment can be used to couple the moir\'{e} mini-bands with other electronic bands. Having seen a rapidly increasing number of different types of moir\'{e} systems~\cite{reviewhe2021moire,reviewlau2022reproducibility,reviewmak2022semiconductor,reviewwang2023moire,reviewwu20242d}, we believe our study opens an interesting venue to investigate new interactions effects in diverse moir\'{e} systems.

\begin{acknowledgments}
The work is financially supported by the National Key R\&D Program of China (2020YFA0309600) and by the University Grants Committee/Research Grant Council of Hong Kong SAR under schemes of Area of Excellence (AoE/P-701/20), CRF (C7037-22G), ECS (27300819), and GRF (17300020, 17300521, and 17309722). K.W. and T.T. acknowledge support from the JSPS KAKENHI (Grant Numbers 21H05233 and 23H02052) and World Premier International Research Center Initiative (WPI), MEXT, Japan.
\end{acknowledgments}
\bibliography{CD}

\appendix

\clearpage
\setlength{\belowcaptionskip}{-0.1cm}
\onecolumngrid
\begin{center}
\section{\MakeTextUppercase{Supplemental Materials for Coulomb drag in graphene/hBN/graphene moir\'{e} heterostructure}}
\end{center}
\vspace{5\baselineskip}
\renewcommand{\thefigure}{S\arabic{figure}}
\renewcommand{\thepage}{S\arabic{page}}
\renewcommand{\bibnumfmt}[1]{[S#1]}
\renewcommand{\citenumfont}[1]{S#1}

\setcounter{figure}{0}
\setcounter{page}{1}
\setcounter{enumiv}{0} 

\section{1. Device fabrication}
\begin{figure}[h]
  \centering
  \includegraphics[width=0.95\textwidth]{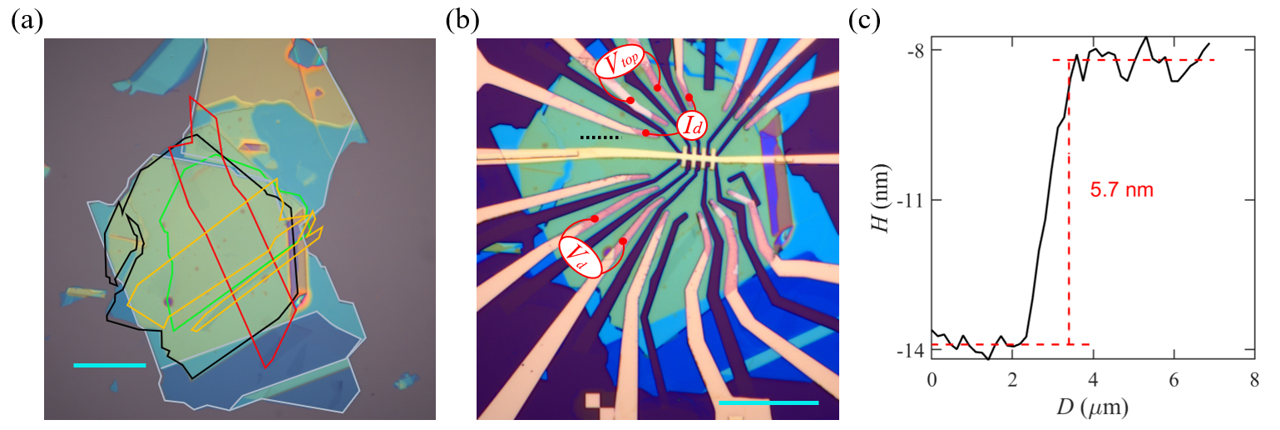}
  \caption{(a,b) Optical image of a typical heterostructure, overlayed with stacking configuration (a). The white, red, green, orange and black sketches denote the edges of the top hBN, top graphene, middle hBN, bottom graphene and bottom hBN respectively. Optical image of the final device with measurement schematics of top drive configuration (b). The scale bar is 20 $\mu$m in (a,b). (c) The thickness of the middle hBN scanned by AFM along the black dotted line in (b).}
  \label{fig:figureS1}
\end{figure}

Graphene and hexagonal Boron Nitride (hBN) flakes are exfoliated from their respective bulk crystals using Scotch tape onto silicon wafers with a 285 nm thick SiO$_2$ layer. The hBN crystals were grown by K. Watanabe and T. Taniguchi and the graphite is from NGS. The monolayer graphene flakes are identified using the optical microscope. The top, middle and bottom hBN thickness is chosen to be $\sim$30nm, 5$\sim$12nm and 30$\sim$50nm, respectively, estimated optically. Once the suitable flakes are identified, top hBN, top graphene, middle hBN, bottom graphene and bottom hBN flakes are picked up using PDMS micro-dome coated with PC film at 100 ℃ [S1], during which we have paid a special attention on the rotational alignment between top (or bottom) graphene and middle hBN to generate moir\'{e} superlattice. Cr/Au (5nm/50nm) electrodes are evaporated on the top of the heterostructures to be used as the top gate using standard electron-beam lithography, electron-beam evaporation, and lift-off followed by the 1D edge contacts to the top and bottom graphene flakes [S2]. Finally, the sandwiched heterostructures are etched into a Hall-bar geometry. The typical assembly and device geometry are shown in Fig.~\ref{fig:figureS1} and Fig.~\ref{fig:figureS1}(c) indicates the thickness of middle hBN ($\sim$5.7 nm) obtained by an atomic force microscope (AFM) scan.

\section{2. Data from another sample}
Fig.~\ref{fig:figureS3} shows the data from another device that exhibits a similar behavior as the device shown in the main text. 
\newpage

\newpage

\begin{figure}[h]
  \centering
  \includegraphics[width=0.95\textwidth]{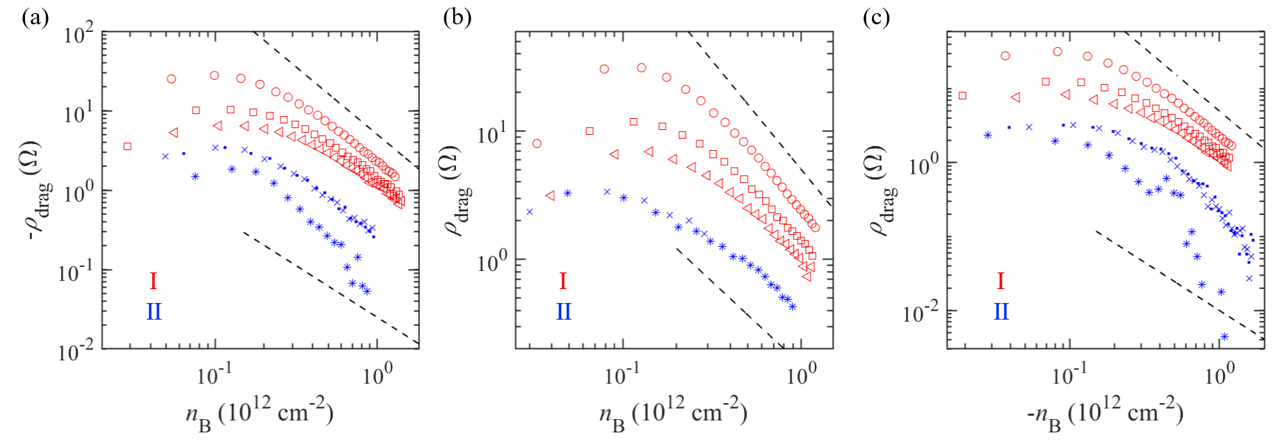}
  \caption{Density dependence of $\rho_{drag}$ in $e-e$ and $e^\ast-e$ section (a), in $h-e$ and $h^\ast-e$ section (b), and in $e-h$ and $e^\ast-h$ section (c). The double-log plots in (a,c) are taken at finite $n_T=n_T^{\ast}=1.2\times10^{11}$ cm$^{-2}$ (circle and asterisk), $3.0\times10^{11}$ cm$^{-2}$ (square and cross), and $4.0\times10^{11}$ cm$^{-2}$ (triangle and dot) with $n_T=n_T^{\ast}$ in (b) the same as in Fig. 2(b) of the main text. The upper and lower dashed lines demonstrate the $n_B^{-1.7}$ and $n_B^{-1.3}$ dependence. Red and blue denote region~\Rmnum{1} and region~\Rmnum{2} respectively. The data shows a similar behaviour as in $h-h$ and $h^\ast-h$ section shown in Fig. 2(b) of the main text.}
  \label{fig:figureS2}
\end{figure}

\newpage

\begin{figure}[h]
  \centering
  \includegraphics[width=0.95\textwidth]{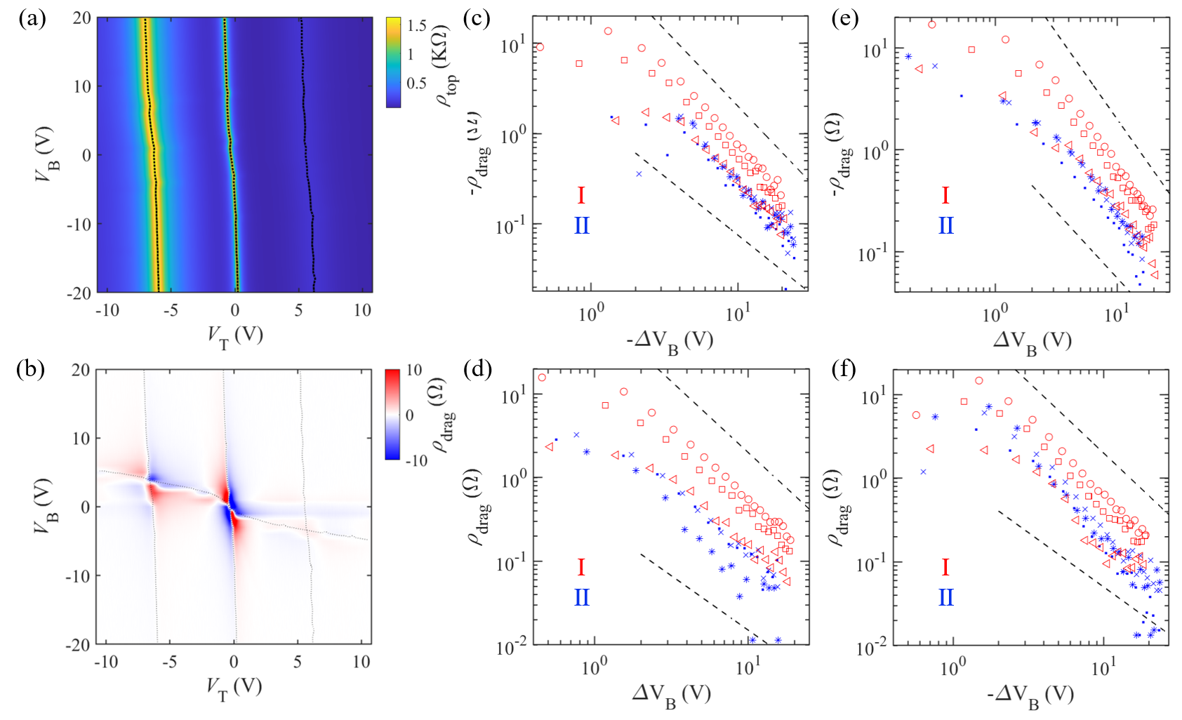}
  \caption{Data from another device presenting moir\'{e} superlattice in top graphene aligned with middle hBN. (a,b) $\rho_{top}$ (a) and $\rho_{drag}$ (b) as a function of $V_T$ and $V_B$ measured at 144 K. The dotted lines are obtained from the local transport measurements. (c-f) Density dependence of $\rho_{drag}$ in four different sections as mentioned earlier, obtained at $\Delta V_T$ = $\pm$0.24 V, $\pm$0.48 V, $\pm$0.96 V. Here, $\Delta V_{T,B}$ = $V_{T,B}$ - $V_{T,B}^0$, where $V^0$ is the gate voltage at the Dirac point. The upper and lower dashed lines scale the ${\Delta V_B}^{-1.7}$ and ${\Delta V_B}^{-1.3}$ separately. Red and blue denote region~\Rmnum{1} and region~\Rmnum{2} respectively. All main features discussed in the main text are reproduced.}
  \label{fig:figureS3}
\end{figure}

\newpage
\clearpage

\section*{Supplemental References}
\begin{itemize}
    \item [[S1]] K. Kim, M. Yankowitz, B. Fallahazad, S. Kang,
H. C. Movva, S. Huang, S. Larentis, C. M. Corbet,
T. Taniguchi, K. Watanabe, et al., Nano Lett. 16, 1989
(2016).
    \item [[S2]] L. Wang, I. Meric, P. Huang, Q. Gao, Y. Gao, H. Tran,
T. Taniguchi, K. Watanabe, L. Campos, D. Muller, et al.,
Science 342, 614 (2013).
\end{itemize}

\end{document}